# The 7th International Conference on the Theory

Diagrams are very wide-ranging and open-ended representations that include sketches, drawings, charts, pictures, 2D and 3D geometric models, and maps. Diagrams are a vital tool in human communication in areas such as art and science, as well as commerce and industry. Increased understanding of how effective diagrams can be generated and used has the potential to produce transformative advances in these areas. Research topics include understanding diagrammatic reasoning in humans; understanding the use of diagrammatic representation for communication; developing techniques for automated diagrammatic reasoning; and designing tools for use of diagrammatic representations.

The goal of the Diagrams 2012 Graduate Symposium is twofold. Firstly, the Symposium will provide senior graduate students and recent graduates with the opportunity to present their research and receive feedback from established researchers who will provide comments on each of the presentations. Secondly, the Symposium will provide students with an opportunity to network with each other as future colleagues. The doctoral student symposium will increase the exposure and visibility of young graduate student researchers in these areas, and train them by providing early input and feedback from senior researchers in the field in an interactive and constructive environment.

## Organising Committee

| | |
|---|---|
| General Chair | Peter Rodgers (University of Kent, UK) |
| Joint Program Chairs | Phil Cox (Dalhousie University, Canada)<br>Beryl Plimmer (University of Auckland, NZ) |
| Tutorials Chair | Gem Stapleton (University of Brighton, UK) |
| Workshops Chair | Nathaniel Miller (University of Northern Colorado, USA) |
| Graduate Symposium Chair | Lisa Best (The University of New Brunswick, Canada) |
| Publicity Chair | Aidan Delaney (University of Brighton, UK) |

# Table of Contents

# Work in Progress: Annotating Gestures in Communication through Graphs

Özge Alaçam, Cengiz Acartürk

Cognitive Science Program, Informatics Institute,

Middle East Technical University, 06800, Ankara/Turkey

{ozge.alacam, acarturk}@metu.edu.tr

**Abstract.** This work in progress study presents a syntactic-level analysis of gestures in communication through annotated line graphs in time domain. As distinct from the gesture analysis at discourse level, the syntactic gesture analysis is based on solely physical properties of the gestures, such as direction, size or position. The results revealed strong relationship between gestures, verbal descriptions and graphical annotations.

## 1 Introduction

Graphs and diagrammatic representations are widely used means of visual communication. In addition to graphical representations, communication is also achieved by the contribution of other modalities, such as gestures. For annotating gestures, one of the mostly preferred methods in the literature is proposed by McNeil and Levy [1]. According to this method, there are four main categories of gestures: beats, deictic, iconic and metaphoric. However, depending on the context, one gesture can belong to more than one category, and this complicates the coding procedure thus increasing the subjectivity. In order to address this issue, McNeill emphasized another method called "Dimensional Framework" [1], which claims that each gesture may have loading for each gesture property. So, a gesture can be multidimensional and the saliency of the dimensions is of the interest. Moreover, adding syntactic level of analysis prior to discourse analysis mentioned above is another method for gesture annotation. At the syntactic level, gestures are analyzed according to their physical properties, such as direction and size. However, gesture and spoken language are distinct forms of communication and these methods are generally engineered for the purpose of gesture-language interaction analysis. On the other hand, the analysis method for gesture–graph interaction should benefit from the shared physical properties based on spatial information carried by both modalities. One of the studies, in which this method (syntactic and discourse level analysis respectively) is employed, was conducted by Acarturk and Alacam [2]. The secondary results of the study indicated that for the evaluation of gesture - graph interaction, the syntactic level seems more informative than the discourse-level analysis that suffers from ambiguity. Regarding this issue, our research interest is to investigate whether a significant relationship can be obtained between the information conveyed in different modalities. The analysis has been performed at a syntactic level based on solely physical properties of gestures.

## 2 Pilot Study

Seven participants (university students or teaching assistants) participated in the study. The experiment was conducted in the native language of all the participants (Turkish). The participants were asked to present single-sentence verbal descriptions of annotated graphs to a hypothetical audience. Participants' spontaneous gestures for 14 annotated population-graphs of bird species were video-recorded. Three types of annotations were used: process annotation, durative state annotation and punctual state annotation (Fig.1).

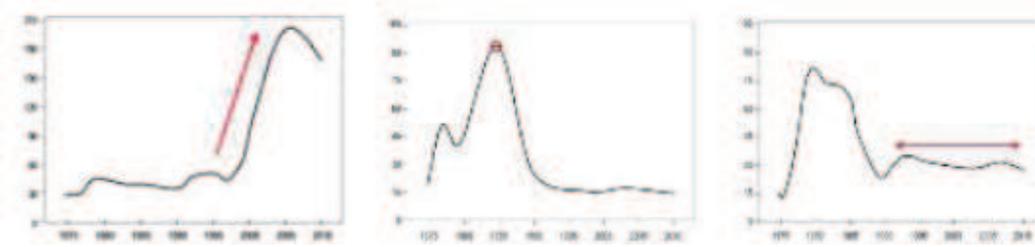

**Fig. 1**. Sample annotated graphs with a process annotation (left), a punctual state annotation (middle), and a durative state annotation (right)

***Syntactic Gesture Annotation.*** The coding scheme based on syntactic features was employed by using The Noldus Observer XT. Each gesture was classified in terms of its directionality: no direction, vertical/diagonal, horizontal, or other. Then the coder identified the following features of each gesture: size, palm direction, speed and start position. In the present study, we focus on directionality of the gestures by leaving the analysis of other features to a further study.

***Spoken Language Transcription.*** The sentences produced by the participants were transcribed and then the parts of the sentences with accompanying gestures were segmented into phrases. After this process, the phrases were classified into five categories: *time*, *state*, *process*, *durative* and *other*. The *time* category covers phrases with temporal information, such as "between 1990 and 2010". The phrases that refer to particular information in the graph, such as a peak, were classified as *state*. The verbs or nouns that point out the change in the population, such as an increase or a decrease formed the *process* category., The *durative* category covers the verbs or nouns that refer a stable duration in population. Finally, the remaining phrases, such as the species of the bird or their location were evaluated under the *other* category.

## 3 Findings

The interactions between the pairs of modalities were analyzed separately and McNemar tests were conducted to investigate the relationships. The syntactic analysis of the gestures based on physical features revealed significant associations with other modalities. Table 1 presents a co-occurrence matrix for three-way interaction between the modalities.

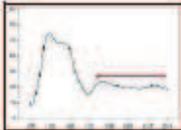

Table 1. Co-occurrence(%) matrix for gesture, speech and graphical annotation

| | | Graphical Annotations | | | | | | | | | |
|---|---|---|---|---|---|---|---|---|---|---|---|
| | | Durative | | | Process | | | | Punctual | | |
| | Word Category: | Time | Process | Durative | Time | Process | States | Durative | Time | Process | States |
| Gesture | No Direction | 27,27 | 4,55 | 9,09 | 20,75 | <5% | 7,55 | - | 20,90 | 10,45 | 37,31 |
| | Vertical/Diagonal | - | <5% | - | - | 32,08 | - | - | - | 5,97 | <5% |
| | Horizontal | - | 9,09 | 40,91 | <5% | <5% | - | <5% | - | - | - |
| | Other | - | <5% | - | <5% | 18,87 | <5% | - | <5% | 8,96 | - |

e.g. co-occurrence of horizontal movement and durative word in accompanying speech is 41% of all produced gestures for the graph, that is annotated with durative state annotation.

## 4 Conclusion

The results point out that durative graphical annotations yield more horizontal gestures and the utterance of the words like "being stable". On the other hand, more vertical/diagonal movements were produced by the participants when describing graphs with process annotations. Additionally, the co-occurrence matrix reveals that verbal descriptions that refer to time or to population, which are accompanied by punctual annotations, are usually not accompanied by directional gestures. These results reveal that syntactic-level analysis of annotations is promising in revealing the interactions between language, gesture and graphical annotations. We state that the annotation method that focuses on syntactic features is not trivial, due to their possible contribution in corpus design regarding gesture modality. In addition, it may also have an important role in decreasing the subjectivity in coding process, since the syntactic features consist of discrete categories that reduce the level of ambiguity for spatial information.

# Classification Tree Diagrams in Health Informatics Applications

Farrukh Arslan

Dept. Electrical and Computer Engineering, Purdue University, West Lafayette, IN, USA
farslan@purdue.edu

**Abstract.** Health informatics deal with the methods used to optimize the acquisition, storage and retrieval of medical data, and classify information in health-care applications. Healthcare analysts are particularly interested in various computer informatics areas such as; knowledge representation from data, anomaly detection, outbreak detection methods and syndromic surveillance applications. Although various parametric and non-parametric approaches are being proposed to classify information from data, classification tree diagrams provide an interactive visualization to analysts as compared to other methods. In this work we discuss application of classification tree diagrams to classify information from medical data in healthcare applications.

**Keywords:** classification tree, anomaly detection, knowledge representation

## 1 Introduction

Medical data in health informatics applications contains patient locations, temporal component (patient's date of visit), syndrome classifications to various categories by ICD codes for time series analysis. Data also comprises age and gender of the individuals and text-field collected patient complaints for outbreak assessment [1]. Domain experts and healthcare analysts are often interested to classify information from medical data for further analysis and exploration. By analyzing extracted information from data, high incidence of events, anomalies or hotspots can be detected and mitigative measures can be taken for prevention.

Most information extraction methods rely on either parametric statistical formulation such as change-point or non-statistical formulation such as decision or classification tree [2]. The decision tree classifiers are constructed for datasets by using the training data over relatively long timeframes. In a decision tree, each internal node involves some test on one or more attributes, while the leaf nodes provide the potential decision information. Traversing the tree from the root to each leaf node, relevant information patterns can be identified and further analyzed by domain experts.

Classification tree diagrams are simple to interpret by domain experts and provide information visualization. Classifications are robust in a sense that there is no assumption made about underlying distribution of data and aprior distribution needs not to be known before building a classification tree diagram.

## 2 Medical Data in Health Informatics

Emergency department data can be used for evaluating algorithms and methods, analyzing multivariate patterns and attribute relationships thereby extracting useful information. Medical data, stored in a relational database is shown in table 1. This data comprises chief complaints which are noted at the emergency department prior to the patient seeing a doctor. Relational medical database also includes patient's home addresses (longitude, latitude) and demographic attributes (age, gender).

**Table 1.** Medical dataset stored in a relational database. Following HIPPA regulations, patient and hospital names have been replaced by ID's. Patient's address is represented by longitude and latitude.

| Patient ID | Hospital ID | Syndrome ID | Date | Chief Complaint | Age | Gender | Latitude | Longitude |
|---|---|---|---|---|---|---|---|---|
| 1 | 0 | 7 | 01/01/06 | R HAND INJURY | 40 | m | 40.86 | -86.88 |
| 2 | 0 | 7 | 01/01/06 | RIGHT HAND INJURY | 21 | m | 40.87 | -86.92 |
| 3 | 0 | 0 | 01/01/06 | NAUSEA WITH VOMITING | 54 | m | 40.8 | -86.97 |
| 4 | 0 | 7 | 01/01/06 | UNKNOWN | 18 | m | 40.81 | -86.87 |
| 5 | 0 | 7 | 01/01/06 | BACK PAIN | 38 | f | 40.81 | -86.94 |
| 6 | 0 | 0 | 01/01/06 | ABDOMINAL PAIN | 89 | f | 40.82 | -86.93 |
| 7 | 0 | 7 | 01/01/06 | TOOTH PAIN | 19 | f | 40.82 | -86.98 |

## 3   Classification Tree Diagram Induction

A classification tree diagram [2] represents a structured binary tree which performs a split test in its internal nodes based on some or more attribute values and predicts decision in its leaf node. Classification trees are built by recursive partitioning. The ultimate goal of classification tree induction algorithm is to end up with the smallest tree and purest leaf nodes i.e. further splitting would not increase information gain. The goal of building a classification tree structure is to have leaf nodes where all the attributes in it are correctly classified, maximizing the information gain.

Each leaf node in a tree is associated with a class. In anomaly detection methods [3], this would indicate if and where an anomaly is present in the dataset while traversing from root. Anomalies that are addressed by classification trees are the patterns in data that do not conform to normal semantic relationships between attributes. Each non-terminal node is associated with one of the attributes that domain rules possess. Each branch is associated with a particular value that the attribute of its parent node can take in dataset. Attribute discrimination is done by calculating Information gain and Entropy.

If an experiment has n possible outcomes then the amount of information *I*, expressed as bits is defined to be entropy, where pi is the prior probability of the i$^{th}$ outcome. The information gain for an attribute 'A' when used with a set of domain rules X is given below.

$$I \equiv -\sum_{i=1}^{n} p_i \log_2 p_i \quad \text{and} \quad Gain(X,A) \equiv I(X) - \sum_{v \in values(A)} \frac{|X_v|}{|X|} I(X_v) \quad (1)$$

Tree diagram resulting from data classification is shown in figure 1. Gender and age attributes shows semantic relationships with complaints/syndromes. For example, elevated blood pressure complaint is related to age factor greater than 35 and elevated blood sugar with age between 11 and 17. Such semantic patterns would be of particular interest for healthcare analysts while exploring records for outbreak assessment. As a future work diagrams can be extended for multi-class knowledge extraction with spatio-temporal patterns spanning over multiple domains.

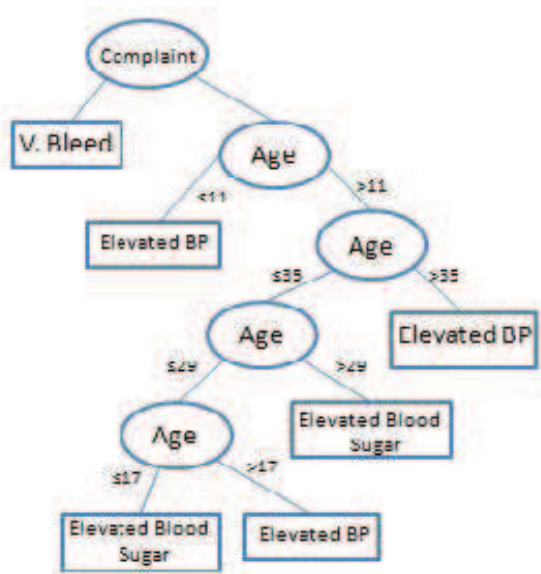

**Fig. 1.** Diagram shows age/gender based attributes classification from medical data. Traversing the tree from the root to each leaf node, semantic patterns can be analyzed.

## 4 Future Work

Classification trees diagrams can be further extended to explore temporal anomalies in time series data, spanning over multiple spatio-temporal domains.

# What Makes an Effective Euler Diagram?

Andrew Blake[1], Gem Stapleton[1], Peter Rodgers[2], and John Howse[1]
[1]University of Brighton, Brighton, UK
{A.L.Blake,G.E.Stapleton,John.Howse}@brighton.ac.uk
[2]University of Kent, Canterbury, UK
P.J.Rodgers@kent.ac.uk

**Abstract.** Euler diagrams can be an effective representation of information when they are both *well-matched* and *well-formed*. However, being well-matched and well-formed alone does not imply effectiveness. Other diagrammatical properties need to be considered. Information visualisation theorists have known for some time that orientation, shape and colour have the potential to effect our interpretation of diagrams. This paper explains why well-matched, well-formed drawing principles are insufficient, illustrates why we should study the orientation, shape and colour of Euler diagrams, and briefly describes our research approach for this endeavor.

## 1 Introduction

Euler diagrams represent set theoretic relationships using interconnected closed curves often drawn using circles or ovals. Curves are labelled, so affording context to the information or data therein. Figure 1 contains three Euler diagrams, all representing the same information, and illustrates there are syntactic choices to be made when using Euler diagrams to visualise data. Each diagram tells us that Course Leaders are a subset of Lecturers, Lecturers are a subset of Academics, and these staff could be Managers. Later in the paper we will discuss the syntactic differences between the diagrams in Figure 1.

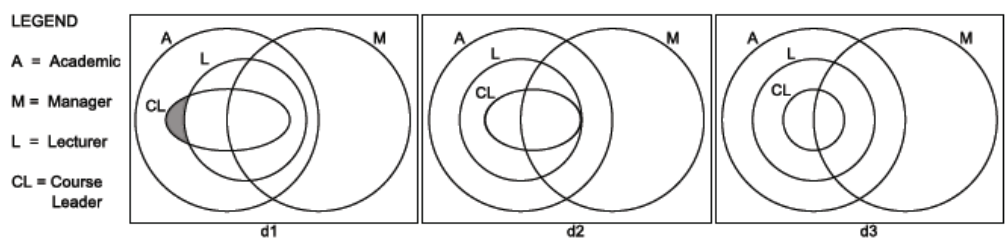

**Fig. 1.** Staff Hierarchy

Euler diagrams are regarded a natural and effective way to depict sets and their relationships. They are found in a wide variety of contexts including architecture, arts and social media. Wilkinson [9] presents a survey of natural science journals and online affiliated content from 2009 observing 72 occurrences of Euler diagrams. To support their use in information visualisation, various researchers have devised automated drawing techniques. SetVisualiser [7] and Vennmaster[1] are examples of software development projects that automatically draw Euler diagrams. It is essential that automatically generated Euler diagrams convey information accurately. Given the importance of accuracy and the variety of contexts in which Euler diagrams are used it is necessary to understand what makes an effective choice of diagram to represent any given data set. Having this understanding will inform the development of these tools and guide users towards drawing effective Euler diagrams.

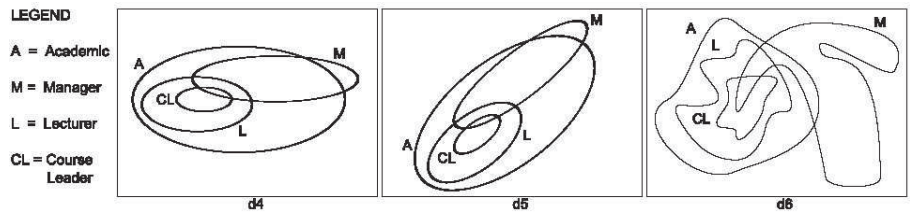

**Fig. 2.** Staff hierarchy

## 2 Effective Euler Diagrams

We will focus on two categories of so-called well-matched and well-formed drawing principles. These are designed to yield effective diagrams where effective means reducing the average time taken to interpret them. Gurr [2], theorising well-matched diagrams, postulates that the most effective diagram is one with structure and property that matches, or closely matches, that which it strives to represent. Well-formedness describes relationships between curves and regions in a diagram. Rodgers et al. [6] empirically test these well-formedness properties observing the extent to which they impact comprehension. Gurr's theory tell us to select well-matched diagrams and Rodgers et al.'s empirical work guides us to select well-formed diagrams.

In figure 1, d1 is neither well-matched or well-formed. It is not well-matched as its shaded region denotes that there are no Course Leaders that are not Lecturers: the set of Course Leaders is contained by the set of Lecturers but CL is not contained by L. It is not wellformed as it has a disconnected zone: the region inside A and L but outside CL comprises two disconnected pieces. The diagram d2 is well-matched but not well-formed. It is not well-formed as it has a disconnected zone, as described earlier, and a brushing point where CL meets L. The diagram d3 is both well-matched and well-formed. It does not exhibit any extraneous properties and the relationship between its curves and regions are neither disconnected or brushing and, therefore, it is regarded as the most effective at conveying information pertaining to staff hierarchy. There are a number of other well-formedness conditions that a diagram can exhibit. One example is concurrency which exists when two or more curve segments follow the same path [6].

## 3 Research Problems

In addition to well-matched and well-formed, there are other diagrammatical properties to consider when ascertaining the effectiveness of a visual representation. For example, perceptual theorists know that we are sensitive to the diagrammatical properties of orientation, shape and colour [8]. Aware of this phenomena, information visualisation theorists manipulate these properties, affecting our interpretation and, thus, comprehension of diagrams [3].

Figure 2 represents the same information as figure 1. Diagrams d3, d4, d5 and d6 are all well-matched and well-formed and, by this property, regarded equally effective. However, there are clear visual differences between them which can be largely attributed to the shape of their curves. Diagram d3 uses circles, diagrams d4 and d5 use ellipses, and diagram d6 uses irregular shapes. Diagrams d4 and d5 are identical but diagram d5 has been rotated -38 degrees. Diagrams d3 to d6 visually illustrate that well-matched well-formed drawing principles alone are too naive in yielding effective Euler diagrams. Conscious that other diagrammatical properties affect our interpretation of diagrams this research is aiming to ascertain the extent to which orientation, shape and colour impact our comprehension of Euler diagrams. For example, which diagram is more effective, d4 or d5? Which shape or combination of shapes are the most effective? The convex and concave attributes of the irregular curves in diagram d6 make them difficult to distinguish. If we colour each curve differently will they be easier to distinguish?

To meet these aims, this research will adopt an empirical approach. Consistent with other researchers who have investigated user comprehension [4, 5], we will record the time taken to answer questions as the primary dependent variable. For example, if orientation impacts on comprehension then we would expect to see, for some diagram, a significant difference between the mean time taken to answer the posed question under one orientation as compared to another orientation. Such a study exploring orientation has already been undertaken. We have yet to perform a detailed analysis of the data but early indications suggest that orientation does not effect our comprehension of Euler diagrams. If true, diagrams d4 and d5 can be said to be equally effective conveying information pertaining to staff hierarchy. Perhaps more significantly, this observation will underpin the work of existing researchers in this community who have already made this assumption [6].

# The Serial Colour (SECO) Profile
# A Colour-Based Form of Graphical Representation

A. Luke MacNeill & Lillian P. Fanjoy
University of New Brunswick
Department of Psychology, Box 5050, Saint John, NB E2L 4L5
{s8me9, d9za0}@unb.ca

**Abstract.** Colour has historically been underemphasized in the graphical representation of data. The current paper proposes a colour-based Serial Colour (SECO) profile, and illustrates using sample data from the Eating Attitudes Test (EAT-26). Typical and atypical examples are illustrated and compared.

**Keywords:** SECO, Serial colour profile, Colour graph

## 1 Introduction

Colour has traditionally been underemphasized in the graphical representation of data. This minimization can be attributed to practical concerns. Colour printing is more difficult to produce and more cost-prohibitive than black-and-white printing. With the advancement and proliferation of computer technology, however, colour graphs are now easier to both create and mass reproduce [1]. The development of new graphical forms that are centered on colour might be beneficial, as they could provide a different perspective on data than one might receive from a line or bar graph.

The current paper proposes a new, colour-based form of graphical representation called a Serial Colour (SECO) profile. With a SECO profile, information from a rating scale questionnaire is tallied and converted into a sequence of proportional colour gradients within a square figure. This is illustrated using sample data from the short version of the Eating Attitudes Test (EAT-26), a 26-item questionnaire that is used to identify individuals with severe eating disturbances such as anorexia nervosa [2]. Global and factor scores are converted into SECO profiles, and typical and atypical examples are compared.

## 2 Methods

Data was obtained for 69 female undergraduate students who completed the EAT-26 questionnaire. For demonstrative purposes, participants with missing data were deleted listwise. This amounted to 5 participants, leaving a pool of 64 participants.

SECO profiles were constructed. First, questionnaire answers were tallied for each participant, so that each person had one score for each answer category: *Always* (indicative of higher scores on the EAT-26), *Usually*, *Often*, *Sometimes*, *Rarely*, and *Never* (the least eating-disordered answer). Question 26 was a reverse-scored question, and this was taken into consideration during the analysis. These tallies were then converted into percentages of the participant's overall answers, which were subsequently represented as proportional areas on a square figure. The two extreme answers were depicted by two colour anchors: red for Always, and blue for Never. Intermediate answers were represented by gradient colours between these two colour anchors.

This procedure was repeated for group scores. In addition to a profile for the total sample, those with a BMI outside of the normal range (18.5 to 25) were isolated. This resulted in two additional group profiles: underweight (BMI below 18.5) and overweight/obese (BMI above 25). Factor scores were calculated for each of these groups, with 13 questions contributing to the dieting factor, six questions contributing to the bulimia and food preoccupation factor, and seven questions contributing to the oral control factor. Profiles were created to illustrate these factor scores.

## 3 Results

Figure 1A shows the SECO profile for the total sample. The blue-skewed profile indicates a greater number of non-disordered answers, with progressively fewer answers for disordered eating. Figure 1B shows the profile for Participant 18, an atypical example. The red-skewed profile indicates a greater amount of eating disordered answers. Figures 1C and 1D show the profiles for two participants who had identical clinical scores on the EAT-26 questionnaire. Upon examination of their profiles, one can see clear distinctions in their answering patterns. Participant 2 (1C) tended to give more moderate answers, with fewer answers at either extreme. Participant 24 (1D) gave a greater proportion of answers at both extremes, with fewer moderate answers.

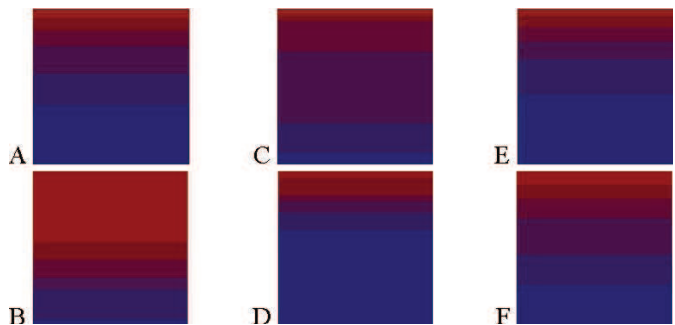

**Fig. 1.** SECO profiles for total sample (Panel A), participant 18 (B), Participant 2 (C), Participant 24 (D), underweight group (E), and overweight/obese group (F).

Panels 1E and 1F show the respective profiles for the underweight and overweight/obese groups. These groups show similar answering patterns to the total sample, with slightly fewer disordered answers from the underweight group, and slightly more disordered answers from the overweight/obese group. Their factor scores show subtleties in responding, however. Dieting subscale scores (Figure 2, top row) showed more disordered responding for the overweight/obese group (2C). Bulimia and food preoccupation scores (Figure 2, middle row) showed less disordered responding for the underweight group (2E). Oral control scores (Figure 2, bottom row) showed more disordered responding for the underweight group (2H).

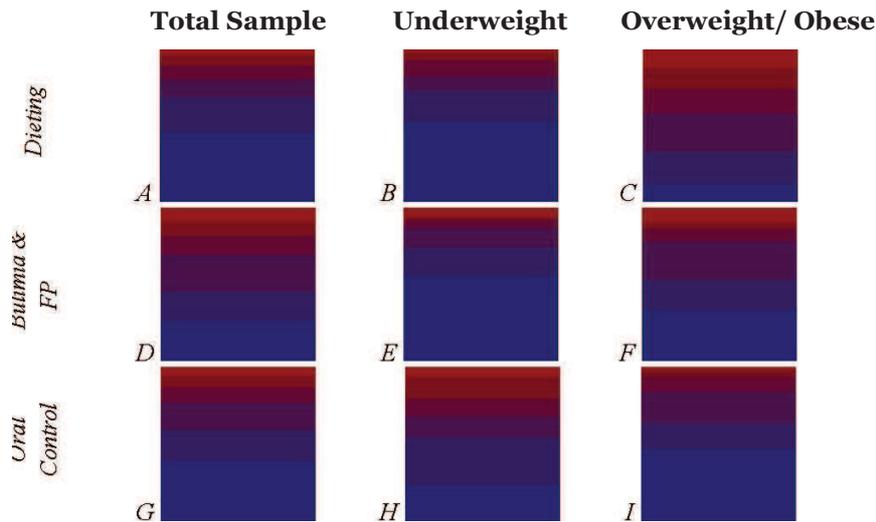

**Fig. 2.** SECO profiles for dieting (Panels A, B, C), bulimia and food preoccupation (Panels D, E, F), and oral control (Panels G, H, I) factor scores; for total sample, underweight group, and overweight/obese group.

## 4 Discussion

SECO profiles are useful for demonstrating variations in rating scale answering patterns. They provide an immediate impression due to their colour dimension. In addition, they illustrate subtleties in questionnaire answering that might not be available from a global numerical score. Complexities could also be added. Questionnaire items could be weighted by importance, and proportional scores could be placed within each segmental area (or directly outside the graph area) for more numerical precision.

## 5 References

1. Wainer, H.: Visual Revelations: Graphical Tales of Fate and Deception from Napoleon Bonaparte to Ross Perot. Copernicus, New York (1997).

2. Garner, D.M., Olmsted, M.P., Bohr, Y., & Garfinkel, P.E.: The Eating Attitudes Test: psychometric features and correlates. Psychological Medicine, 12, 871-878 (1982)

# Towards a Visual Compositional Relational Programming Methodology

Gorkem Pacacı, Andreas Hamfelt
Department of Informatics and Media Uppsala University

**Abstract.** We present a new visual programming method, based on Combilog, a compositional relational programming language. In this paper we focus on the compositional aspect of Combilog, the make operator, visually implementing it via a modification of Higraph diagrams, in an attempt to overcome the obscurity and complexity in the textual representation of this operator.

## 1 Introduction

Recent interest in parallel programming has put declarative programming methodologies such as functional programming and relational programming even more in focus thanks to their natural aptitude for modularity and parallelization [5]. Combilog, as a point of departure for our study, is a compositional relational programming language. The compositional nature of Combilog comes from the make operator, presented in detail in the following sections. In this study, we focus on eliminating the textual complexity that arises from this generalised projection operator using diagrams.

In the following subsections, we present the programming language, Combilog, and the steps to represent Combilog programs with Harel's Higraph [5]. In the sections that follow, we explain how we combine the make operator and Higraphs, as a first step to form a visual language.

### 1.1 Combilog

Programming languages offer ways of producing new components out of available ones. In Functional Programming, one example is higher-order functions. In Logic Programming, there are also languages like λProlog which allow writing higher-order predicates.

Combilog follows a similar path for logic programming, applying Quine's Predicate-Functor Logics developed and updated in [7] and [6]. Instead, higher-level predicate functors are used to permute, replicate or exclude argument-places [3]. Combilog generalizes these functors into a single operator called make that can satisfy the same requirements for modification of argument-places, as explained in [2]. It also introduces two quasi higher-order predicates as recursion operators: foldr and foldl. [2] and [4] explain in detail, how make, foldr and foldl operators, together with n-ary conjunction and disjunction, and a few pre-defined ground predicates provide a programming model. It is important to note that in this paper we focus on visualization of only one of the apparatus in Combilog, the make operator. Together with boolean expressions, make operator builds the composite structure of Combilog. Other operators like folds, and any other ground predicates are components or 'operands', while the 'make', 'and', and 'or' are the composition operators.

The make operator A definition of the make operator from [4] is:

make[µ1, ..., µm](Q)(X$\mu_1$, ..., X$\mu_m$) ← Q(X1, ..., Xn)

Notice that since m and n have no restrictions in magnitude in relation to each other, and since the expression allows arbitrary ordering of argument-places in the resulting predicate, it is possible to remove or reorder arguments of, and introduce arguments to, a given predicate. An example of application of the make operator from [4], given that the pre-defined pairing predicate cons is defined as cons(U,V,[U|V]):

head ← make[3, 1](cons)
in Combilog, translates to Prolog code:
head(L, X) ← cons(X, ,L)

## 2 The visual representation

### 2.1 Make operator and Higraphs

To narrow the problem, we only attempt to visualize Combilog expressions that include make and and operators, since the difference between visualizing 'and' and 'or' operators are minor rather than structural. Let us refer to this type of Combilog expressions as conjunction of makes, and the predicate identifier that this expression is assigned to as the new predicate. Here we define the modifications we execute upon the original Higraph diagram.

1. Every atomic set in a Higraph represents an argument in one of the predicates in conjunction. If two or more arguments of different predicates are bound by conjunction, they are represented by a single atomic set.

2. Every non-atomic set in a Higraph represents a predicate. Directed acyclic edges must be used to order the arguments of a predicate. We call these ordered edges argument-orders. For unary predicates, no edges exist.

3. Following from Combilog's variable-free form [2], we remove set identifiers from atomic sets, replacing their visual representations with solid filled circles. From this point on, we call these argument-places.

4. The argument-places which are contained in the new predicate become black, the others become gray. Every argument-order is also assigned a colour.

5. Last, we remove contours for all non-atomic sets except those represent unary predicates and the new predicate. We place predicate names on the first edge of argument-orders. Removing contours increases readability by reducing visual interference and entanglement.

# 3 Conclusion

As mentioned earlier, make expressions in Combilog may become obscure as the complexity increases. One simple instance of this issue is provided below: siblings(X, Y) ← parent(Z, X), parent(Z, Y), X\= Y

The Combilog equivalent can be written as: siblings ← make[1, 2](and(and(make[2, 3, 1](parent), make[3, 2, 1](parent)), make[1, 2, 3](ineq)))

With the suggested method, we can visualize Combilog code above with the Higraph diagram below. Predicates ineq, parent[1] and parent[2] are represented with colours red, green and blue.

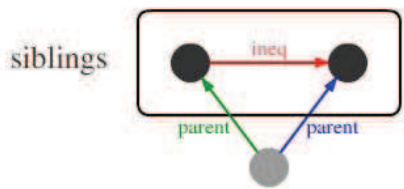

We have presented a visual model of representing make expressions of Combilog, using Higraphs. We believe this method to be a viable and efficient ground for designing a new visual logic language with compositional relational nature.

# A Sketch Recogniser as a Virtual Machine

Alistair G. Stead

The University of Cambridge, Computer Laboratory, Cambridge, UK,
ags46@cam.ac.uk,
http://www.alistairstead.me

**Abstract.** A vast number of people regularly produce content in the form of whiteboard sketches, paper-based notes or even art, which is difficult to transfer into a digital equivalent. Sketch recognition systems could provide a solution, but have proven to give poor results in unconstrained domains. Multi-domain recognisers address this problem by focusing on a single-domain at a time, while being applicable to several. The extension of such a system however, requires programming experience. I aim to use techniques drawn from end-user programming (EUP) and machine learning to allow users without programming experience to construct disposable sketch recognisers. I describe the motivations, and conclude by on arguing that sketch recognition is increasingly becoming an EUP problem, rather than a purely technical challenge.

**Keywords:** Sketch Recognition, End User Programming, Programming By Example

## 1 Sketch Recognisers

Sketch recognition has long been a difficult problem: general-purpose recognisers inevitably encounter issues of precision, ambiguity and performance due to complexity. Furthermore, in most informal cases, the intention of a sketcher affects the semantics of the sketched object. One solution is to focus on the domains of interest, limiting the set of possible classications and making the problem analogous to those we are already capable of solving. These single-domain recognisers can be used successfully for industrial applications, or in specialised circumstances, however the majority of users are unable to benefit. Plimmers InkKit [8] enables multi-domain recognition, but requires an interpreter for each domain, and therefore cannot be extended by non-programmers. Rata.SSR [4] can be trained on several examples to produce an accurate recogniser. Issues with this system include the requirement for several training examples, and the lack of semantic description of sub-components and therefore of the sketch.

## 2 Extending Sketch Grammars

Sketch grammars describe a hierarchy of sketch items in which complex components are described in terms of lower level items. There have been efforts to use these formalisms for sketch recognition [3] [7]. LADDER [7] demonstrates how this strategy can improve recognition rates and improve sketch understanding.

Domain Builder [6] allows instructors to extend the LADDER grammar. The user is also able to program an extension class to link the recogniser to other systems. Although custom recognisers can be constructed in a reasonable time, this approach has several drawbacks that are common to programming systems, and are highlighted when considering the Cognitive Dimensions of Notations framework [5]: the user must think in an abstract manner, prematurely commit to object descriptions, and do some hard mental operations to be able to describe the object in terms of its subcomponents.

## 3 Sketch Recognition by End User Programming

The ubiquity of modern computing devices, such as smartphones, is likely to increase demand for sketch recognition. I conducted an initial user study to consider how and why sketches are transformed into digital artefacts. Results suggest that users take pictures of collaborative sketches for future reference. Images are then used to aid creation of a more formal digital artefact (such as a document). They are rarely translated directly, due to edits and additions, or the need to improve sketches that might be perceived to be unfinished [2]. A study in the software engineering domain [10] has also found that sketches are regularly photographed, scanned, reused, and archived.

Despite potential demand, creating and extending recognition tools usually requires programming experience. The proposed system extends sketch grammars using concepts from Programming by Example (PBE). Programming in this context, describes user actions ranging from parameter tweaking to scripting, rather than conventional programming. PBE is a form of end-user programming, in which a user gives an observing system examples of a repetitive task. In this case, given one or more sketch items with the same semantics, the system should infer a description and constraints to extend the sketch grammar, thus enabling accurate recognition of the object in the future. Sketch-recognition by example is expected to significantly reduce recogniser construction time. Furthermore, it would enable usage scenarios that were previously not possible.

## 4 Operational Transforms

Several systems have linked recognition to programming; BrightBoard [9] is an example of predefined notation affecting execution of an observing system. More recent systems have used sketches to guide computation [1]. This functionality has potential to be powerful in particular usage scenarios, although end users are again constrained by lack of programming experience. An inference-based sketch recognition system could allow users to attach functionality to items. This would extend recognition by treating items as notation in a programming language.

The use of operational transforms (OTs) in recognition systems could present unwelcome issues. The level of formality in sketching may also be increased by drawing pre-defined items. I will conduct a user study to determine whether changes in sketching formality affect user behaviour in collaborative scenarios. The study is intended to give an indication as to whether the inclusion of OT functionality would be advantageous, and to enable a thorough system evaluation.

## 5 Conclusion

I have presented a vision for a system that enables quick construction of sophisti-cated sketch recognisers that target a given domain. The intention is to empower end-users without programming experience, while maintaining the high recogni-tion rates of single-domain recognisers. Users would describe semantics of new items, aiding overall sketch understanding and interpretation. I have also pre-sented operational transforms as a system extension. OT Logic would be defined using a simple visual language. On item recognition, the specified logic would be executed, either immediately or at some point in the future. I have also described a user study devised to discover the implications of formality on sketching. I hope to conduct further studies after implementation to investigate real world usage scenarios and the ways in which OTs can be used. I hope to contribute to our understanding of how diagrams and formal/informal sketches are used.

# Enhanced vector diagrams for illustrating elementary mechanics

Peter J. Vivian
Coventry University with Blackpool & the Fylde College

**Abstract.**  A phenomenological study of mechanics textbooks over the last 200 years has resulted in the development of an enhanced form of vector diagram. This diagram has an expanded symbol notation and more developed geometric syntax based on illustration exemplars from the past. Action research with a group of graphic design students supports the thesis that the system will help students acquire a deeper conceptual understanding of mechanics and facilitate their application of mathematics in this domain.

**Keywords:** line metaphor, vector diagrams, symbol-based writing, dynamics

## 1 Introduction

This research is about the graphic metaphor of a line used to reify forces and motions (vectors) that is no longer evident in the illustrations of mechanics text books, why this is a problem and what we can do about it. Fig. 1. The metaphoric basis of vector addition There exists a mathematics problem in engineering education today (Kent & Noss 2003) and this research is important because it has found a new factor and suggests a way of overcoming its effects. This factor concerns the use of the line to reify forces and motions. This line metaphor enables the geometric analogy that is used to create a mathematical model for the analysis of vectors. See figure 1. The position of the line represents the point of application; the orientation of the line represents the line of action and the length of the line is used to denote the magnitude. The graphic manipulation of these lines reflects reality and can be modeled with mathematics.

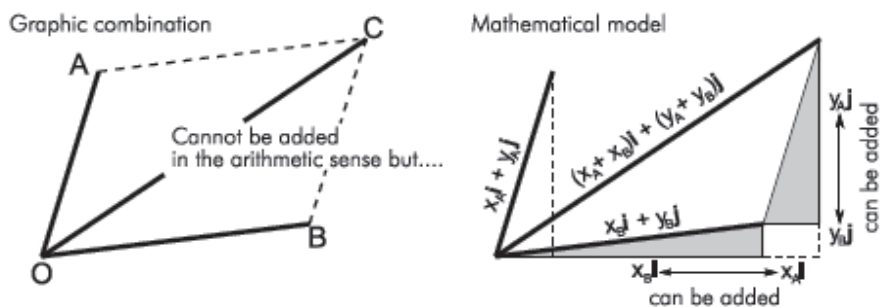

**Fig. 1.** The metaphoric basis of vector addition

There exists a mathematics problem in engineering education today (Kent & Noss 2003) and this research is important because it has found a new factor and suggests a way of overcoming its effects. This factor concerns the use of the line to reify forces and motions. This line metaphor enables the geometric analogy that is used to create a mathematical model for the analysis of vectors. See figure 1. The position of the line represents the point of application; the orientation of the line represents the line of action and the length of the line is used to denote the magnitude. The graphic manipulation of these lines reflects reality and can be modeled with mathematics.

Visual metaphors are often used to create symbols used in diagrams but once their signification is understood, there is no need for the metaphor (Blackwell 1998:1). However, vector diagrams are within a class of diagrams that use the line metaphor as part of their syntax. The metaphor needs to be understood in order to be able to construct and interpret the diagrams. Current literature demonstrates that all graphic communication can be perceived as a picture or read like text (Harris 1986:122) and this explains why domain experts derive more significance from diagrams than other viewers. This research contributes to this debate by revealing the role of the graphic line metaphor in creating the scriptorial aspect of the diagram syntax and that vector diagrams are more appropriately defined as semasiographic (symbol-based) writing systems (Sampson 1985:29).

## 2 Phenomenological research

Phenomenological research into the illustrations in mechanics text books over the last 200 years has revealed the pedagogic paradigm shift away from geometric explanation to mathematical modeling and analysis. One change is that illustrations are not used in direct explanation of mechanical principles but are used simply to decode algebraic signs. The consequential loss of metaphoric significance is given as a possible aggravation of the mathematics problem in engineering education today. Saussure's (1995:97) model of a linguistic sign is used to argue that mathematics lacks the semiotic ability to apply metaphor and hence explain elementary concepts. This shortcoming also explains why mathematics needs a geometric contrivance to analyze dynamics. A symbol-based writing system, with an expanded symbol set and an enhanced syntax derived from the geometric manipulations of the past, is proposed that increases the scriptorial aspect of this contrivance and hence its ability to communicate conceptually. See figure 2.

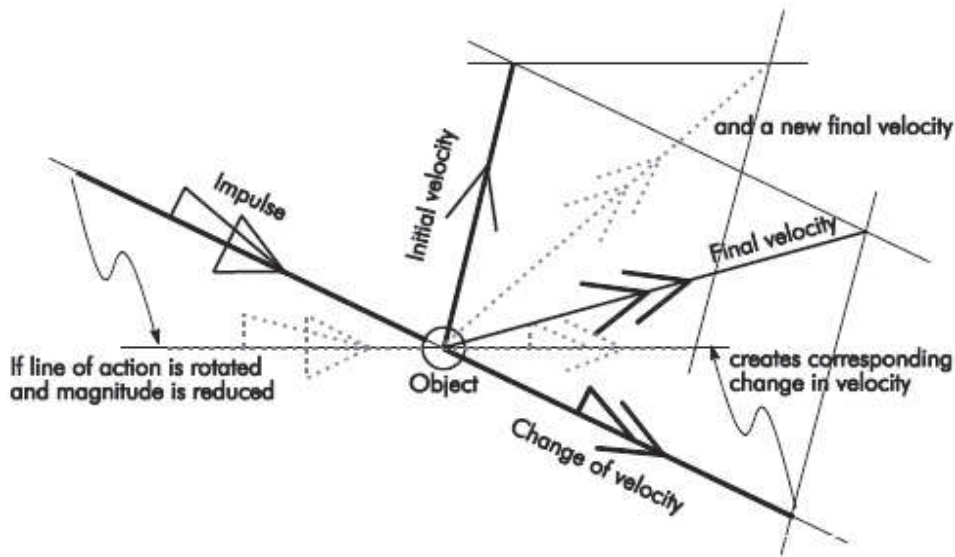

**Fig. 2.** Graphic representation of Newton's 2nd Law of Motion

Figure 2 shows this notation being used to describe Newton's second law of motion i.e. the equivalent of **F**t = m(**v**2-**v**1). The impulse (**F**t) symbol creates the change in velocity (**v**2-**v**1) symbol of the same length and they both share the same line of action. So when one is changed the other has to change in a corresponding manner. The parallelogram of velocities is used to determine the effect on the velocity of the object (m).

## 3 Action research

An experimental study based on the Cheng (2002:701) study of AVOW diagrams was used to simulate the pedagogic experience of explain, practice and test. The concept of acceleration was selected as it is a known threshold concept in the domain of vector dynamics. First year graphic design students participated in the study and two iterations were achieved. Despite being ignorant in the domain of dynamics, these participants were able to assimilate in varying degrees the generic (vector) nature of acceleration. The research concludes that this system has the potential to reinforce the metaphoric basis of vector algebra and hence contribute to the resolution of the mathematics problem in the domain of mechanics.

## 4 Recommendations for further research

As well as continuing to develop the system for dynamics, the writing system can be expanded to include graphic statics and axial vectors, which would involve expanding the symbol notation and the spatial syntax. There is also the issue of how to accommodate scalar properties, e.g. mass & Newton's 2nd law, energy. This research has demonstrated that the system could be used to teach elementary dynamics but it has not demonstrated that the effort is worthwhile in terms of the learner's subsequent experience. It is proposed that a short course in elementary mechanics be devised that covers the principle concepts of statics and dynamics using this writing system and the subsequent performance of the participants be investigated. The course would also facilitate the continued development of the system.